# Hydration Water Dynamics and Instigation of Protein Structural Relaxation

Daniela Russo[1], Greg Hura[2], Teresa Head-Gordon[1,2]

[1]Department of Bioengineering and [2]Graduate Group in Biophysics, University of California, Berkeley CA 94720

Until a critical hydration level is reached, proteins do not function[1]. Restoration of some enzymatic activity is observed in partially hydrated protein powders to suggest that the dynamical and structural properties of the surface water are intimately connected to protein stability and function[3-15]. This critical level of hydration is analogous to a similar lack of protein function observed for temperatures below a temperature range of 180-220K that also is connected to the dynamics of protein surface water[1,2]. Many elegant studies using both experiment and simulation have contributed important information about protein hydration structure and timescales[3,4,6-14,16-19].

The molecular mechanism of the solvent motion that is required to instigate the protein structural relaxation above a critical hydration level or transition temperature has yet to be determined. In this work we use quasi-elastic neutron scattering (QENS) and molecular dynamics simulation to investigate hydration water dynamics near a greatly simplified protein surface. We consider the hydration water dynamics near the completely deuterated N-acetyl-leucine-methylamide (NALMA) solute, a hydrophobic amino acid side chain attached to a polar blocked polypeptide backbone, as a function of concentration between 0.5M-2.0M, under ambient conditions. We note that a folded protein's surface is roughly equally distributed between hydrophobic and hydrophilic domains, whose lengthscales are on the order of a few water diameters[20], and which justify our study of hydration dynamics of the simple NALMA system.

The systematic study of the NALMA water hydration dynamics provides an important dissection of hydration dynamics near inhomogeneous protein surfaces, with implications for supercooled liquids, protein folding and function, and protein-protein interfaces. In this Communication, we focus our results of hydration dynamics near a model protein surface on the issue of how enzymatic activity is restored once a critical hydration level is reached, and provide a hypothesis for the molecular mechanism of the solvent motion that is required to trigger protein structural relaxation when above the hydration transition.

The QENS experiment was performed at the NIST Center for Neutron Research, using the disk chopper time of flight spectrometer (DCS). In order to separate the translational and rotational components in the spectra, two sets of experiments were carried out using different incident neutron wavelengths of 7.5Å and 5.5Å to give two different time resolutions. The spectra were corrected for the sample holder contribution and normalized using the vanadium standard. The resulting data were analyzed with DAVE programs (http://www.ncnr.nist.gov/dave/). The AMBER force field[21] and SPCE water model[22] were used for modelling the NALMA solute and water, respectively. For the analysis of the water dynamics in the NALMA aqueous solutions, we performed simulations of a dispersed solute configuration consistent with our previous structural analysis[23], where we had primarily focused on the structural organization of these peptide solutions and their connection to protein folding[23,24]. Further details of the QENS experiment and molecular dynamics simulations are reported elsewhere[25].

The QENS data arising from translational water dynamics of these biological solutions are analyzed in a first approximation with a jump diffusion model. At the highest solute concentrations, corresponding to a (shared) single layer of water, the hydration dynamics is significantly suppressed and characterized by a long residential time and a slow diffusion coefficient, similar to supercooled water at –10ºC, and a rotational relaxation time of about 2.2ps. The analysis of the more dilute concentration solutions, corresponding to approximately 2-3 hydration layers of water per solute, has been performed taking in account the results of the 2.0M solution as a model of the first hydration shell. Subtracting the first hydration layer based on the 2.0M spectra, the translational diffusion dynamics is still suppressed, although the rotational relaxation time and residential time are converged to bulk-water values.

The experimental "elastic incoherent structure factor" (EISF) can be interpreted as a measure of the fraction of "*immobile*" or localized hydrogen rotational dynamics that are faster or slower than our experimental resolution of 1-5.5ps[26,27]. The EISF shows significant evolution between 0.5M-2.0M; the EISF for the 0.5M solution measures 37% of immobile hydrogens, whereas only 17% of the protons are not observed for the 2.0M concentrations. Figure 1 presents the hydration water EISF variation as a function of NALMA concentration.

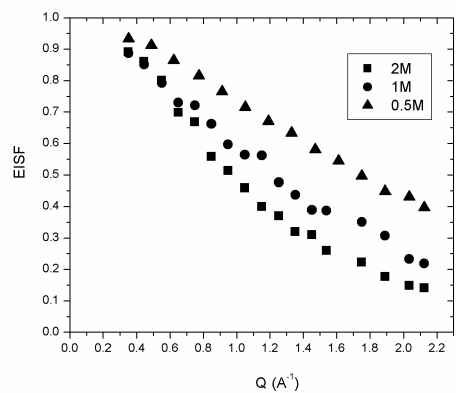

**Figure 1** EISF of hydration water plotted versus $Q^2$, for 0.5M, 1M and 2.0M NALMA concentration.

This seemingly puzzling result was analyzed by molecular dynamics simulation, in which the residence times of water molecules near the NALMA solute were monitored at the hydrophobic side chain and the polar backbone separately. To analyze the EISF results, we evaluated the rotational dynamics of water molecules that maintained a distance of 4.0Å or less from the branching carbon center of the hydrophobic side chain, and within 4.0Å of one of the backbone carboxyl oxygens of the NALMA molecule, as well as an average residence time of that subset of water molecules.

Figure 2 presents the orientation autocorrelation function, $P_2(t)$, of water molecules with these residence times. The 0.5M $P_2(t)$ data are well fit with two exponentials, and show populations with very slow rotational timescales (~6-8 ps) and fast rotational timescales (~1ps). The total 2.0M $P_2(t)$ data is best fit with one exponential, with underlying slow rotational timescales (~4-5ps) near the hydrophilic site, and faster rotational timescales (~2 ps) near the hydrophobic site. A stretched exponential model also provided a good fit to the autocorrelation function of the 2M data, with a β–exponent value between 0.4-0.6. This complementary analysis confirms that the 2M NALMA concentration shows a distribution of rotational time scales.

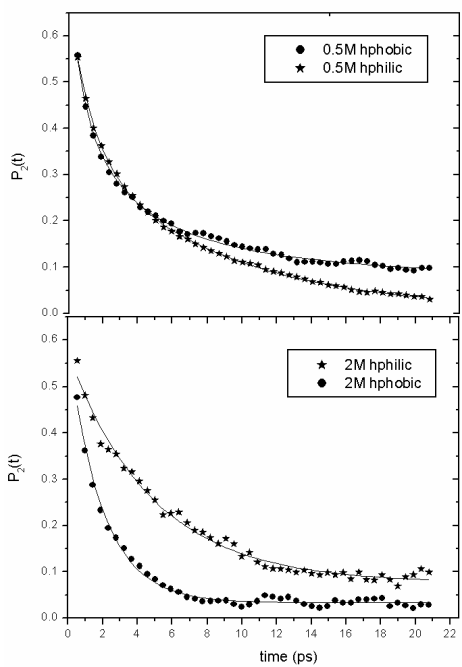

*Figure 2* The rotational dynamics were analyzed using the orientational autocorrelation function: $P_2(t)=<0.5[3\cos^2\theta(t) -1]>$ where $\theta(t)$ measures the angle between the dipole vector of the water molecule at times t and 0. We evaluated the rotational dynamics of water molecules that maintained a distance of 4.0Å or less from the branching carbon of the hydrophobic side chain, and within 4.0Å of the backbone carboxyl oxygen of the NALMA molecule.

The MD simulations provide an interpretation of the EISF results, in which the 0.5M solute concentration shows the presence of long rotational relaxation times near both the hydrophilic and hydrophobic side chains, while the 2.0M solute concentration shows more water protons near the hydrophobic side chain whose dynamics become faster, and therefore resolvable by the QENS experiment.

The MD simulations also measured first layer water residence times near hydrophobic and hydrophilic sites. We find that at 0.5M the water residence times are largely equal near both sites: ~8.5-9.0ps; however at 2.0M we find that the water residence times are very different between the two sites: ~3.5-4.0ps near the hydrophobic side chain, whereas it is ~10.0-10.5ps near the hydrophilic site. Qualitatively we attribute the higher percentage of localized hydrogens seen at 0.5M as arising from the better ability to form more idealized hydrogen-bonded networks around the hydrophobic side chain. By contrast the reduced levels of immobile hydrogens at higher NALMA concentrations results in a more strained-non-optimal network that breaks more easily to permit faster motions.

We focus our QENS and MD results of hydration dynamics near a model protein surface on the issue of how enzymatic activity is restored once a critical hydration level is reached, and provide a hypothesis for the molecular mechanism of the solvent motion that is required to trigger protein structural relaxation when above the hydration transition. Below the critical hydration level, water dynamics near hydrophobic sites is much faster and incommensurate with that near the hydrophilic sites; it is *too fast* to effectively solvate the hydrophobic side chains, and the hydrogen-bonded water network across the protein surface is dynamically unstable. At a sufficient level of hydration, the hydration dynamics become spatially homogeneous, with restoration of a water network that can support hydrophobic hydration over the surface with time scales that are *slow* enough to couple to protein conformational transitions to realize the structural plasticity necessary for protein function.

**Acknowledgment** We gratefully acknowledge the support of NIH GM65239-01. This work utilized facilities supported in part by the National Science Foundation under Agreement No. DMR-0086210. We acknowledge the support of the National Institute of Standards and Technology, U.S. Department of Commerce, in providing the neutron research facilities used in this work. The authors acknowledge Dr. John Copley for assisting on the DCS spectrometer.

(1) Rupley, J. A.; Careri, G. *Advances in Protein Chemistry* **1991**, *41*, 37-172.
(2) Doster, W.; Settles, M. In *Hydration Processes in Biology; Les Houches Lectures*; Bellissent-Funel, M. C., Teixeira, J., Eds.; IOS Press, 1998.
(3) Bellissent-Funel, M. C. *Journal of Molecular Liquids* **2000**, *84*, 39-52.
(4) Bizzarri, A. R.; Cannistraro, S. *Journal of Physical Chemistry B* **2002**, *106*, 6617-6633.
(5) Careri, G.; Peyrard, M. *Cellular & Molecular Biology* **2001**, *47*, 745-756.
(6) Dellerue, S.; Bellissent-Funel, M. C. *Chemical Physics* **2000**, *258*, 315-325.
(7) Denisov, V. P.; Halle, B. *Faraday Discussions* **1996**, 227-244.
(8) Denisov, V. P.; Jonsson, B. H.; Halle, B. *Nature Structural Biology* **1999**, *6*, 253-260.
(9) Halle, B.; Denisov, V. P. *Biophysical Journal* **1995**, *69*, 242-249.
(10) Mattos, C. *Trends in Biochemical Sciences* **2002**, *27*, 203-208.
(11) Otting, G. *Progress in Nuclear Magnetic Resonance Spectroscopy* **1997**, *31*, 259-285.
(12) Tarek, M.; Tobias, D. J. *Journal of the American Chemical Society* **1999**, *121*, 9740-9741.
(13) Tarek, M.; Tobias, D. J. *Biophysical Journal* **2000**, *79*, 3244-3257.
(14) Tarek, M.; Tobias, D. J. *Physical Review Letters* **2002**, *8813*, 8101.
(15) Zanotti, J. M.; Bellissent-Funel, M. C.; Parello, J. *Biophysical Journal* **1999**, *76*, 2390-2411.
(16) Svergun, D. I.; Richard, S.; Koch, M. H. J.; Sayers, Z.; Kuprin, S.; Zaccai, G. *Proceedings of the National Academy of Sciences of the United States of America* **1998**, *95*, 2267-2272.
(17) Diehl, M.; Doster, W.; Petry, W.; Schober, H. *Biophysical Journal* **1997**, *73*, 2726-2732.
(18) Marchi, M.; Sterpone, F.; Ceccarelli, M. *Journal of the American Chemical Society* **2002**, *124*, 6787-6791.
(19) Russo, D.; Baglioni, P.; Peroni, E.; Teixeira, *Chemical Physics,* **2003** 292/2-3, 235-245.
(20) Janin, J. *Structure with Folding & Design* **1999**, *7*, R277-R279.
(21) Cornell, W. D.; Cieplak, P.; Bayly, C. I.; Gould, I. R.; Merz, K. M.; Ferguson, D. M.; Spellmeyer, D. C.; Fox, T.; Caldwell, J. W.; Kollman, P. A. *J. Am. Chem. Soc.* **1995**, *117*, 5179-5197.
(22) Berendsen, H. J.; Grigera, J. R.; Straatsma, T. P. *J. Phys. Chem.* **1987**, *91*, 6269-6271.
(23) Sorenson, J. M.; Hura, G.; Soper, A. K.; Pertsemlidis, A.; Head-Gordon, T. *Journal of Physical Chemistry B* **1999**, *103*, 5413-5426.
(24) Hura, G.; Sorenson, J. M.; Glaeser, R. M.; Head-Gordon, T. *Perspectives in Drug Discovery & Design* **1999**, *17*, 97-118.
(25) Russo, D.; Hura, G.; Head-Gordon, T. *Biophysical Journal* **2003**, *submitted*.
(26) Bellissent-Funel, M. C.; Teixeira, J.; Bradley, K. F.; Chen, S. H. *Journal de Physique I* **1992**, *2*, 995.
(27) Zanotti, J. M.; Bellissentfunel, M. C.; Parello, J. *Physica B* **1997**, *234*, 228-230.

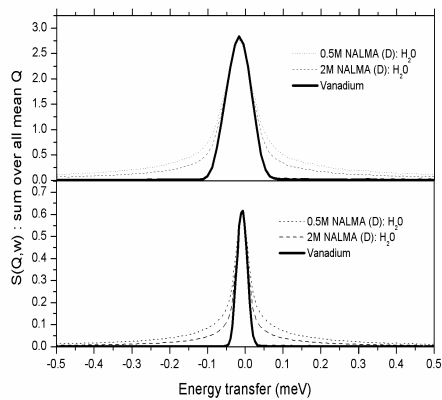

ABSTRACT FOR WEB PUBLICATION The evolution of water dynamics from dilute to very high concentration solutions of a prototypical hydrophobic amino acid with its polar backbone, N-acetyl-leucine-methylamide (NALMA), is studied by quasi-elastic neutron scattering and molecular dynamics simulation for the completely deuterated leucine monomer. We observe several unexpected features in the dynamics of these biological solutions under ambient conditions. We find analogy between the dynamics of supercooled liquids and the hydration dynamics of the most concentrated solutions, with translational diffusion and rotational motions that correspond to supercooled water at −10ºC. Molecular dynamics analysis shows spatially varying dynamics at high concentration which becomes spatially homogeneous at more dilute concentrations. These results allow us to provide a hypothesis for the molecular mechanism of the solvent motion that is required to trigger protein structural relaxation and the onset of protein function when above a critical hydration level. In essence, a sufficient level of hydration is required for the water dynamics to become spatially homogeneous, with restoration of a water network that can support hydrophobic hydration over the protein surface with time scales that are *slow* enough to couple to protein conformational transitions to realize the structural plasticity necessary for protein function.